\documentclass[sn-mathphys-num]{sn-jnl}% Math and Physical Sciences Numbered Reference Style 
\usepackage{graphicx}%
\usepackage{multirow}%
\usepackage{amsmath,amssymb,amsfonts}%
\usepackage{amsthm}%
\usepackage{mathrsfs}%
\usepackage[title]{appendix}%
\usepackage{xcolor}%
\usepackage{textcomp}%
\usepackage{manyfoot}%
\usepackage{booktabs}%
\usepackage{algorithm}%
\usepackage{algorithmicx}%
\usepackage{algpseudocode}%
\usepackage{listings}%

\newcommand{\be}{\begin{equation}} 
\newcommand{\ee}{\end{equation}} 
\begin{document}

\title[The Tolman-Ehrenfest criterion of thermal equilibrium in 
scalar-tensor gravity]{The Tolman-Ehrenfest criterion of thermal equilibrium in 
scalar-tensor gravity}


\author[1]{\fnm{Numa} \sur{Karolinski}}\email{numa.karolinski@mail.mcgill.ca}

\author*[2]{\fnm{Valerio} \sur{Faraoni}}\email{vfaraoni@ubishops.ca}
\equalcont{These authors contributed equally to this work.}

%\author[2]{\fnm{Third} \sur{Author}}\email{iiiauthor@gmail.com}

\affil[1]{\orgdiv{Physics Department}, \orgname{McGill University}, 
\orgaddress{\street{3600 Rue University}, \city{Montreal}, 
\postcode{H3A~2T8}, \state{Quebec}, \country{Canada}}}

\affil*[2]{\orgdiv{Department of Physics \& Astronomy}, \orgname{Bishop's 
University}, 
\orgaddress{\street{2600 College Street}, \city{Sherbrooke}, 
\postcode{J1M~1Z7}, \state{Quebec}, \country{Canada}}}


\abstract{The Tolman-Ehrenfest criterion for the thermal equilibrium of a 
fluid at rest in a static general-relativistic geometry is generalized to 
scalar-tensor gravity. Surprisingly, the gravitational scalar field, which 
fixes the strength of the effective gravitational coupling, does not play 
a role in determining thermal equilibrium. As a result, heat does not sink 
more in a gravitational field where gravity is stronger.}

\keywords{Scalar-tensor gravity, Thermal equilibrium, Tolman-Ehrenfest 
criterion}



\maketitle

\section{Introduction} 
\label{sec:1} 

Thermal physics in curved spacetime is rather intriguing. In the early 
days of general relativity (GR) Tolman, Ehrenfest 
\cite{tolman1928extension,Tolman:1930zza,Tolman:1930ona}, and then Eckart 
\cite{Eckart:1940zz,Eckart:1940te} discussed the thermal equilibrium of 
fluids and heat 
conduction in the relativistic context. The concept of thermal equilibrium 
is not as general, or useful, in GR as it is in pre-relativistic physics, 
or even in special relativity. In fact, when spacetime becomes curved and 
dynamical, a thermal system (e.g., a fluid), is shaken around by the 
time-varying geometry described by the spacetime metric $g_{ab}$, spatial 
temperature gradients and fluid acceleration arise, heat flows, and 
thermal 
equilibrium becomes impossible unless the time scale of microscopic 
processes thermalizing this fluid is much smaller than the time scale of 
variation of the geometry. Sometimes this adiabatic situation occurs, for 
example when particle physics reactions thermalize the coupled 
photon-baryon fluids in the early universe \cite{kolb1990early}, but rapid 
thermalization cannot be guaranteed to occur in general.

Here we revisit the Tolman-Ehrenfest criterion for thermal equilibrium in 
GR 
\cite{tolman1928extension,Tolman:1930zza,Tolman:1930ona,klein1949thermodynamical} and extend it 
to scalar-tensor gravity. There is now a huge literature on scalar-tensor 
gravity, starting in the 1960s with the Jordan-Brans-Dicke theory 
\cite{Brans:1961sx} and its generalizations 
\cite{Bergmann:1968ve,Nordtvedt:1968qs,Wagoner:1970vr,Nordtvedt:1970uv},  
continuing with the 
interest in string and dilaton gravity 
\cite{Green:1987sp,Polchinski:1998rq,Polchinski:1998rr} 
and with a revival of scalar-tensor gravity in the past two 
decades. While, historically, the original motivation for Brans-Dicke 
gravity arose in the context of Mach's principle \cite{Brans:1961sx}, it 
was soon realized that virtually any attempt to quantum-correct GR 
modifies it by introducing curvature corrections to the Einstein-Hilbert 
action (for example, in Starobinski inflation \cite{Starobinsky:1980te} 
which is currently the scenario favoured by observations 
\cite{Planck:2015sxf,WMAP:2012fli}), higher order terms in the field 
equations, 
or extra degrees of freedom such as scalar fields. To wit, in Starobinski 
inflation quadratic corrections are added to the Ricci scalar 
Lagrangian of GR modifying it   to $f(R)=R+\alpha 
R^2$. The 
low-energy limit of the bosonic 
string theory, the simplest string theory, produces an $\omega=-1$ 
Brans-Dicke theory (where $\omega$ is the Brans-Dicke coupling) 
\cite{Callan:1985ia,Fradkin:1985ys}. Other quadratic quantum corrections 
to the curvature produce fourth order field equations instead of the 
second order Einstein equations \cite{Stelle:1976gc,Stelle:1977ry}.

The more recent interest in scalar-tensor gravity is motivated by the 
possibility of explaining the current acceleration of the universe without 
an {\em ad hoc} dark energy, mostly employing $f(R)$ theories which are 
ultimately scalar-tensor ones (see 
\cite{Sotiriou:2008rp,DeFelice:2010aj,Nojiri:2010wj} for reviews).  The 
interest continued in the past decade 
with the rediscovery of Horndeski gravity \cite{Horndeski:1974wa} and its 
extension to Degenerate Higher Order Scalar-Tensor (DHOST) theories 
\cite{Deffayet:2009wt,Deffayet:2009mn,Deffayet:2011gz,Gleyzes:2014dya,Gleyzes:2014qga,Langlois:2015cwa,Langlois:2015skt,BenAchour:2016cay,Crisostomi:2016czh,BenAchour:2016fzp,Crisostomi:2017aim,Faraoni:2023hin}, see 
\cite{Langlois:2018dxi,Langlois:2017mdk,Creminelli:2018xsv,Langlois:2017dyl,Langlois:2018dxi,Kobayashi:2011nu}  
for reviews.

In spite of this literature, the now old Tolman-Ehrenfest criterion for 
thermal equilibrium has not been extended to scalar-tensor gravity, even 
in the simplest scenarios.  Here we fill this gap, beginning with the 
introduction of this criterion and its more modern derivation 
\cite{Misner:1973prb,Santiago:2018kds,Santiago:2018jeu,Santiago:2019aem}. 
We follow the 
notation of Ref.~\cite{Wald:1984rg}: 
$g$ is the determinant of the spacetime metric $g_{ab}$, $\nabla_a$ is its  
covariant derivative operator, $R_{ab}$ is the Ricci tensor, while its 
trace 
$R\equiv g^{ab} R_{ab}$ is the Ricci scalar.

Consider a static spacetime endowed with a timelike Killing vector $k^a$ 
and a test fluid at rest in the frame adapted to this time symmetry, i.e., 
the fluid four-velocity is $u^a=k^a/\sqrt{ -k_ck^c}$. Fixing $u^a$ 
determines the $3+1$ splitting into the time direction tangent to $u^a$ 
and $k^a$ and the 3-space with Riemannian metric \be h_{ab} = g_{ab} +u_a 
u_b \,, \ee which is a purely spatial tensor according to the fluid 
observers, $h_{ab}u^a =h_{ab}u^b=0$. In a coordinate system $\left( 
t,x^i \right)$ adapted to the time symmetry, the line element reads 
\begin{equation} ds^2 = g_{00}(x^k)\, dt^2 + g_{ij}(x^k)\, d x^i\, dx^j \, 
\quad (i,j,k=1,2,3)\,.
 \label{line_element_static_spacetime} \end{equation} Static observers 
have four-velocities parallel to the timelike Killing vector which, in 
these coordinates, has components $k^{\mu} = \left( k^0 , 0, 0, 0 
\right)$.  A test fluid at rest with respect to this static observer has a 
position-dependent equilibrium temperature 
$\mathcal{T} =  \mathcal{T}(x^k)$ satisfying the Tolman-Ehrenfest 
criterion 
\cite{tolman1928extension,Tolman:1930zza,Tolman:1930ona} (see also 
\cite{klein1949thermodynamical,Santiago:2018kds,Santiago:2019aem,Santiago:2018jeu}) 
\begin{equation}
    \mathcal{T}\sqrt{-g_{00}} = \mathcal{T}_0\,, 
    \label{tolman_ehrenfest_criterion_gr}
\end{equation} 
where ${\cal T}_0$ is a constant. Contrary to 
pre-relativistic physical intuition, a fluid in thermal equilibrium in a 
non-uniform gravitational field necessarily has a spatial temperature 
gradient. This gradient expresses the fact that heat is a form of 
mass-energy and sinks in a gravitational field. Therefore, regions where 
gravity is stronger are hotter, in a way described by the gravitational 
shift factor $g_{00}$ 
\cite{Santiago:2018kds,Santiago:2019aem,Santiago:2018jeu}. Here thermal 
equilibrium is defined as the vanishing 
of the heat flux density $q^a = 0$.

The corresponding criterion for the equilibrium of particles with respect 
to diffusion in a static spacetime was formulated by Klein simply by 
replacing the temperature ${\cal T}$ with the chemical potential $\mu$ 
\cite{klein1949thermodynamical}.

Although completely unrelated to gravity, the Tolman-Ehrenhest criterion 
is also of interest for thermal transport in materials. In fact, by 
viewing thermal transport as the linear response of a material to a 
temperature gradient, Luttinger described this phenomenon by introducing a 
counter-balancing weak gravitational field that restores thermal 
equilibrium in the presence of a temperature gradient 
\cite{Luttinger:1964zz}.

The Tolman-Ehrenfest temperature is related to the Hawking temperature of 
a black hole seen by an observer at asymptotic infinity 
\cite{Birrell:1982ix,Wald:1999xu,Giddings:2015uzr}. 
There is also interest in applying the Tolman-Ehrenfest criterion to 
neutron stars \cite{Laskos-Patkos:2022lgy,Kim:2022qlc,Li:2022url} and in 
simultaneous heat conduction and particle diffusion 
\cite{Lima:2019brf,Kim:2021kou}, as well as in the combined equilibrium in 
Weyl-integrable geometries \cite{Lima:2021ccv}. Recent work on the 
Tolman-Ehrenfest criterion has highlighted the severe limitations inherent 
in its extension to stationary but non-static geometries 
\cite{Santiago:2018kds,Santiago:2018jeu,Santiago:2019aem}. It is however, 
possible to extend this criterion to conformally static spacetimes, 
provided that the microphysics ensures that local thermal equilibrium is 
maintained, which is of interest in cosmology and for the Hawking 
radiation of dynamical black holes \cite{Faraoni:2023gqg}.

Back to scalar-tensor gravity: in these theories the strength of the 
effective gravitational coupling is $
    G_\mathrm{eff} = 1 / \phi $, where $\phi$ is the gravitational 
Brans-Dicke-like scalar field. Thus,  one would expect, in 
scalar-tensor gravity, the 
Tolman-Ehrenfest criterion to be modified to include 
$\phi$ in 
such a 
way that, in regions of space where $\phi$ is smaller (i.e., 
$G_\mathrm{eff} $ is larger and gravity is stronger), the fluid 
temperature is higher. However, 
the Tolman-Ehrenfest criterion remains unchanged in both Jordan frame and 
Einstein frame scalar-tensor gravity.

To be more precise, assume that the fluid subject to our investigation is 
described by the dissipative stress-energy tensor \be
    T_{ab} = \rho u_a u_b + P h_{ab} + \pi_{ab} + q_a u_b + q_b u_a\,, 
    \label{stress_energy_tensor_fluid}
\ee where $\rho=T_{ab}u^a u^b $ is the fluid energy density, $u^a 
=k^a/\sqrt{-k_c k^c}$ is its four-velocity, $P = h_{ab} T^{ab}/3 $ is its 
isotropic pressure, $\pi_{ab}= {h_a}^c {h_b}^d T_{cd}- Ph_{ab} $ is the 
anisotropic stress tensor, and $q_a = -{h_a}^b u^c T_{bc} $ is the heat 
flux density \cite{Eckart:1940te}. In GR, Eckart \cite{Eckart:1940zz} 
generalized 
the phenomenological Fourier law describing heat conduction to the 
relativistic domain, as expressed by the constitutive relation of the 
fluid~(\ref{stress_energy_tensor_fluid}) \be
    q_a = -\mathcal{K} h_{ab}\big(\nabla^b\mathcal{T} + 
\mathcal{T}{\dot{u}}^b\big)\,, \label{relativistic_fourier_law} \ee where 
$\mathcal{T}$ and $\mathcal{K}$ are the fluid temperature and thermal 
conductivity, respectively \cite{Eckart:1940te}.  The only significant 
deviation of Eq.~(\ref{relativistic_fourier_law}) from the 
non-relativistic Fourier law is the inertial term (the second term in the 
right-hand side brackets) contributed by the fluid's four-acceleration 
$\dot{u}^a \equiv u^c\nabla_c u^a$.
 
It was shown by Buchdahl \cite{Buchdahl49} that the four-acceleration of a 
particle in a static spacetime is given by \be a^b \equiv \dot{u}^b = 
\nabla^b\,\ln \left( \sqrt{-k^d\,k_d} \, \right)  
\label{four_acceleration_general} \ee or, in coordinates adapted to the 
time symmetry (in which the line element has the 
form~(\ref{line_element_static_spacetime})), \be
    a^b = \nabla^b\,\ln \sqrt{-g_{00}}\,. \label{four_acceleration_static} 
\ee Therefore, in a static spacetime and in these coordinates, the 
condition of thermal equilibrium $q^a=0$ for this static test fluid reads, 
using (\ref{relativistic_fourier_law}) \begin{eqnarray} q_a &=& 
-\mathcal{K} h_{ab} \left( \nabla^b {\cal T} + \mathcal{T}{\dot{u} }^b 
\right)  \nonumber\\
&&\nonumber\\
&=& -\mathcal{KT}h_{ab}\left[ \nabla^b \left( \ln \mathcal{T} \right) +
\dot{u}^b \right] \nonumber\\
&&\nonumber\\
&=& -{\cal KT} h_{ab} \nabla^b \left[
\ln \left( \mathcal{T}\sqrt{-g_{00}} \right) \right] = 0 \,. 
\label{eckart_thermal_equilibrium} \end{eqnarray}

Since the projection of the gradient $\nabla^b \left[ \ln \left( 
\mathcal{T}\sqrt{-g_{00}} \right)\right]$ onto the 3-space orthogonal to 
$u^a$ vanishes, this gradient is parallel to $u^b$, i.e., 
$\mathcal{T}\sqrt{-g_{00}}$ depends only on time. However, since ${\cal 
T}$ and $g_{00}$ are time-independent because of staticity, there 
exists an integration constant ${\cal T}_0$ such that 
Eq.~(\ref{tolman_ehrenfest_criterion_gr}) holds. Let us examine now 
scalar-tensor geometries.

\section{Scalar-tensor gravity} 
\label{sec:2} 

In the Jordan frame of scalar-tensor gravity, the scalar field $\phi$ 
couples explicitly to the Ricci scalar $R$, as described by the action 
\cite{Brans:1961sx,Bergmann:1968ve,Nordtvedt:1968qs,Wagoner:1970vr,Nordtvedt:1970uv} 
\begin{eqnarray}
    S_\mathrm{ST} &=& \frac{1}{16\pi} \int d^4x\sqrt{-g}\Big[\phi R - 
\frac{\omega(\phi)}{\phi}\,\nabla^a\phi \nabla_a\phi-V(\phi)\Big] 
\nonumber\\
&&\nonumber\\
&\, & + 
S_\mathrm{(m)}   = S_\mathrm{(grav)}+ S_\mathrm{(m)}\,,
\label{jordan_frame_scalar_tensor_action} 
\end{eqnarray} 
where 
$\omega(\phi)$ is the Brans-Dicke coupling, $V(\phi)$ is a potential for 
the Brans-Dicke field $\phi$, while $S_\mathrm{(m)}$ is the matter part of 
the action. Varying the Jordan frame 
action~(\ref{jordan_frame_scalar_tensor_action}) with respect to $g^{ab}$ 
yields the field equations 
\begin{equation}
    G_{ab}= \frac{8\pi}{\phi} \, T^\mathrm{(m)}_{ab} +T^{(\phi)}_{ab} \,, 
\end{equation} which can be read as effective Einstein equations where 
$T^\mathrm{(m)}_{ab}$ is the matter stress-energy tensor and 
\begin{eqnarray} 
T^{(\phi)}_{ab} &=& \frac{\omega}{\phi^2} \left( \nabla_a 
\phi \nabla_b \phi -\frac{1}{2} \, g_{ab} \nabla_c \phi \nabla^c \phi 
\right) \nonumber\\
&&\nonumber\\
&\, & +\frac{1}{\phi} \left( \nabla_a \nabla_b \phi
- g_{ab} \Box \phi \right) -\frac{V}{2\phi}\, g_{ab} 
\end{eqnarray} 
is an effective stress-energy tensor of the 
form~(\ref{stress_energy_tensor_fluid}) \cite{Faraoni:2018qdr}. The 
variation of the action~(\ref{jordan_frame_scalar_tensor_action}) with 
respect to $\phi$ gives 
\be 
\Box \phi = \frac{1}{2\omega+3} \left( 
  \frac{8\pi T^{(m)} }{\phi} + \phi \, \frac{d V}{d\phi}
-2V -\frac{d\omega}{d\phi} \, \nabla^c \phi \nabla_c \phi \right) \,, 
\ee 
 where $\Box \equiv g^{ab} \nabla_a \nabla_b $ is the curved space 
 d'Alembertian and $T^\mathrm{( m)} \equiv g^{ab} T^\mathrm{( m)}_{ab}$.

The conformal transformation \begin{equation} g_{ab} \to \tilde{g}_{ab} = 
\Omega^2\,g_{ab} = \phi\,g_{ab}\,
    \label{conformal_transformation} \end{equation} and the scalar field 
redefinition $\phi \to \tilde{\phi}$ given in differential form by \be 
d\tilde{\phi} = \sqrt{ \frac{2\omega+3}{16\pi} } \, \frac{d\phi}{\phi} \ee 
bring the Jordan frame action~(\ref{jordan_frame_scalar_tensor_action}) 
into its Einstein frame version, in which the gravitational sector of the 
theory looks like the Einstein-Hilbert action 
\begin{equation}
    S_\mathrm{(grav)} = \int d^4x \sqrt{-\tilde{g}}\left[ 
\frac{\tilde{R}}{16\pi}  - \frac{1}{2} \,\tilde{g}^{ab} \tilde{\nabla}_a 
\tilde{\phi} \tilde{\nabla_b} \tilde{\phi} -U \left( \tilde{\phi} \right) 
\right] 
\ee 
where a tilde denotes quantities constructed with 
$\tilde{g}_{ab}$, and 
\be U \left( \tilde{\phi} \right) = \frac{ V \left[ 
\phi \left( \tilde{\phi} \right)  \right] }{ \phi^2 \left( \tilde{\phi} 
\right)}\,. 
\ee
 
In the Einstein frame, the equation of timelike geodesics is 
modified to \cite{Dicke:1961gz}
\be
\tilde{ \dot{u}}^a \equiv \tilde{u}^b \tilde{\nabla}_b 
\tilde{u}^a =   \sqrt{ \frac{4\pi}{2\omega+3} } \, 
\tilde{\nabla}^a \tilde{\phi} \,.
\ee
Using $ \tilde{\nabla}_a
\tilde{\phi} =\sqrt{  \frac{2\omega+3}{16\pi}} \, \nabla_a \ln \phi$, 
this corrected geodesic equation reads
\be
\tilde{ \dot{u}}^a =  \tilde{\nabla}^a \left( \ln \sqrt{\phi}  \right) 
\,.\label{modified-geodesic-eq}
\ee
Massive test particles subject only to gravity feel both the metric and 
the gravitational scalar field $\phi$ (or $\tilde{\phi}$) and deviate from 
geodesics.

The decomposition~(\ref{stress_energy_tensor_fluid}) of the stress-energy 
tensor applies to any symmetric two-index tensor \cite{Faraoni:2023hwu}, 
thus the stress-energy tensor $\tilde{T}_{ab}$ of the Einstein frame fluid 
can be decomposed as 
\be
    \tilde{T}_{ab} = \tilde{\rho} \, \tilde{u}_a \tilde{u}_b + \tilde{P} 
\tilde{h}_{ab} + \tilde{\pi}_{ab}
+ \tilde{q}_a \tilde{u}_b + \tilde{q}_b \tilde{u}_a\,,
    \label{conformal_stress_energy_tensor_fluid} \ee without extra 
assumptions on the conformal fluid quantities $\tilde{\rho}$, $\tilde{P}$, 
$\tilde{\pi}_{ab}$ and $\tilde{q}_a$.

The physical implications of conformal transformations, including the 
scaling of dimensional physical quantities and of their units were 
discussed long ago by Dicke \cite{Dicke:1961gz}. According to 
\cite{Dicke:1961gz}, only the ratio of a dimensional quantity to its unit 
(that scales the same way with powers of $\Omega$) is physically 
measurable, therefore physics in the rescaled geometry $\tilde{g}_{ab}$ is 
equivalent to the same physics in the non-rescaled geometry $g_{ab}$, 
provided that ratios of quantities are considered instead of the 
quantities themselves. This is easier said than done and Dicke's point of 
view, in particular with regard to the physical equivalence of Jordan and 
Einstein frames, has been the subject of a heated debate that continues to 
these days (see \cite{Faraoni:2006fx} for a review). However, it is hard 
to 
argue with the fact that, when measuring dimensional quantities one 
measures only a ratio of a quantity to its unit. The complications and the 
disagreements arise when derivatives of these quantities, or their 
combinations, come into play. However, to discuss 
the Tolman-Ehrenfest 
criterion one needs only the temperature, which is an exceptionally simple 
situation and no such complication is involved.

Dicke showed that, under the conformal 
scaling~(\ref{conformal_transformation}), masses scale as $ \tilde{m} = 
\Omega^{-1}\,m $ while times and lengths scale as $ \Delta \tilde{\ell} 
\sim 
\Delta \tilde{t} \sim \Omega$. Since the speed of light {\em in vacuo} $c$ 
is a 
ratio 
of space and time, it is conformally invariant, and since energy has 
dimensions $[Mc^2]$, it scales like a mass 
\cite{Dicke:1961gz,Faraoni:2006fx}. The product $k_B {\cal T}$ (where 
$k_B$ is the Boltzmann constant) is dimensionally a mass and scales as 
$k_B \tilde{{\cal T}}\sim \Omega^{-1}$. Since the Boltzmann constant $k_B$ 
is a 
true constant, then $ \tilde{{\cal T}} \sim \Omega^{-1}$ (which is in 
agreement 
with our Eq.~(\ref{temperature_conformal_relationship}) below).

\section{Thermal equilibrium in scalar-tensor gravity} 
\label{sec:3} 

As in Einstein gravity, thermal equilibrium of a static test fluid in a 
static spacetime is defined as the absence of heat fluxes, $q^a=0$.  We 
discuss the Jordan frame and the Einstein frame and show how the two 
descriptions give the same result.

\subsection{Jordan frame}

In the Jordan frame, assume:

\begin{itemize}

\item a static spacetime with line 
element~(\ref{line_element_static_spacetime}) in coordinates adapted to 
the time symmetry, in which the timelike Killing vector defines the time 
direction, $k^a=\left( \partial /\partial t \right)^a$;

\item a static scalar field $\phi=\phi\left( x^i \right)$;

\item a test fluid described by the stress-energy 
tensor~(\ref{stress_energy_tensor_fluid}), with four-velocity parallel to 
the timelike Killing vector $k^a$ and satisfying the Eckart constitutive 
relation~(\ref{relativistic_fourier_law}). This fluid is assumed to be at 
rest in the static frame.

\end{itemize}

Then, any observer comoving with this fluid will see no heat flow, $q^a=0$, 
and necessarily thermal equilibrium.

The assumption that the Jordan frame fluid satisfies the Eckart 
constitutive relation is crucial, but its legitimacy is not clear {\em a 
priori}. While Eckart's relation seems natural since it expresses the 
response of the fluid to temperature gradients, the fluid is now also in 
thermal equilibrium with the scalar field $\phi$, thus things could 
change. We will justify this assumption in the next subsection by mapping 
a more familiar Einstein frame result back to the Jordan frame and showing 
that Eckart's relation holds in the Jordan frame if and only if it holds 
in the Einstein frame. With this {\em caveat}, let's proceed. The 
derivation of the Tolman-Ehrenfest criterion $T\sqrt{-g_{00}}=$~const. 
then proceeds the same way as in GR.

\subsection{Einstein frame}

We now derive the Tolman-Ehrenfest criterion in the 
Einstein conformal frame, 
where the theory formally looks like GR and then we use it to 
justify the Eckart constitutive relation and the Tolman-Ehrenfest 
criterion in Jordan frame scalar-tensor gravity. 

The key idea of the calculation is that, since we know the criterion for 
thermal equilibrium in GR, and that Einstein frame 
gravity formally 
looks like GR, we can begin our discussion in this conformal frame 
and then map 
the Tolman criterion back to the Jordan frame, which is the one most 
commonly used to formulate physics in scalar-tensor gravity. As 
explained in Appendix~\ref{Appendix:A}, while the Einstein frame fluid 
couples explicitly to the scalar $\tilde{\phi}$, this coupling is 
automatically included in Eckart's constitutive relation and does not need 
to be supplemented by extra heat flux density terms introduced by hand.

Since $\Omega=\sqrt{\phi \left( x^i\right)}$ is static, $ 
u^b\nabla_b\Omega = 0$, and the conformal geometry with metric 
$\tilde{g}_{ab}=\phi \, g_{ab} $ is also static.  An exception is given by 
stealth solutions of the scalar-tensor field equations in which $g_{ab}= 
\eta_{ab}$ (the Minkowski metric), $\phi$ does not gravitate, but it is 
non-trivial and time-dependent 
\cite{Ayon-Beato:2005yoq,Robinson:2006ib,Ayon-Beato:2004nzi,Demir:2006ed,Hassaine:2006gz,Takahashi:2020hso,Faraoni:2022jyd}.  
This situation is discussed in Sec.~\ref{sec:4}.

Since $\tilde{g}_{ab}=\Omega^2 \, g_{ab}$ is static, the Buchdahl relation 
\cite{Buchdahl49} holds in the rescaled world, 
\be
\tilde{a}^b = \tilde{\nabla}^b\left( \ln \sqrt{-\tilde{g}_{00}} \right)\,. 
    \label{conformal_four_acceleration_assumption}
\ee \
This relation can also be proved directly in the Einstein frame (see 
Appendix~\ref{Appendix:A}).

Under a conformal transformation $\tilde{T}_{ab}$ maintains its 
dissipative fluid form~(\ref{conformal_stress_energy_tensor_fluid}), which 
is common to any symmetric two-index tensor 
regardless of its nature \cite{Faraoni:2023hwu}. 
Furthermore, the $\big(3+1\big)$ splitting of spacetime into space and 
time does not change because the time direction $\tilde{u}^c = u^c / 
\Omega 
$ of the observers comoving with the fluid in the tilded world is parallel 
to the time direction $u^c$ of the observers comoving with the Jordan 
frame fluid, while $\tilde{h}_{ab}=\Omega^2 \, h_{ab}$ is just a rescaling 
of the Riemannian 3-metric $h_{ab}$.  Then, $g_{ab} = -u_au_b + h_{ab}$ 
has the parallel decomposition \be \tilde{g}_{ab} = \Omega^2g_{ab} = 
\Omega^2\big(-u_au_b + h_{ab}\big)=-\tilde{u}_a\tilde{u}_b + 
\tilde{h}_{ab} \ee in the tilded world, as follows from the fact that 
$\tilde{u}_a = \Omega \, u_a$ and $\tilde{h}_{ab}=\Omega^2 \, h_{ab}$. 
Since $h_{ab}u^a = h_{ab}u^b = 0$, it is also $ \tilde{h}_{ab} \tilde{u}^a 
= \tilde{h}_{ab} \tilde{u}^b = 0$ and purely spatial vectors or tensors 
with respect to $g_{ab}$ remain purely spatial with respect to 
$\tilde{g}_{ab}$. In particular the Einstein frame counterpart of the heat 
flux density, which satisfies $\tilde{q}^c =0 $, is purely spatial, i.e., 
$\tilde{q}_c \tilde{u}^c = 0$.

As in the Jordan frame, in the Einstein conformal frame the spacetime is 
static and the test fluid is at rest in the static frame of the metric 
$\tilde{g}_{ab}$. Thermal equilibrium is defined again as the absence of 
heat flux, i.e., $\tilde{q}_a = 0$.  If one is willing to assume that the 
constitutive relation~(\ref{relativistic_fourier_law}) is valid in the 
Jordan frame, it should also hold for the conformal fluid in the Einstein 
frame. Indeed, under a conformal transformation $g_{ab}\to 
\Omega^2 g_{ab}$, the heat flux density transforms according to 
$\tilde{q}_a=\Omega^{-3} q_a$ (see Eq.~(\ref{macheczz}) below).  
As a consequence, the notion of thermal equilibrium does not  
depend on the conformal frame: $q^a$ vanishes in the Jordan frame if and 
only if $\tilde{q}^a$ vanishes in the Einstein frame.  Indeed, it would be   
worrisome if thermal equilibrium depended on the conformal frame.

 Then, at thermal equilibrium, it must be 
\begin{eqnarray} 
\tilde{q}_a &=& -\tilde{\mathcal{K}}\tilde{h}_{ab}\left( \tilde{\nabla}^b 
\tilde{\mathcal{T}} + \tilde{\mathcal{T}}{\dot{\tilde{u}}}^b \right) 
\nonumber\\
&&\nonumber\\
&=& -\tilde{\mathcal{K}}\tilde{\mathcal{T}}\tilde{h}_{ab}\left[
\tilde{\nabla}^b\big(\ln \tilde{\mathcal{T}}\big) + \dot{\tilde{u}}^b 
\right] \nonumber\\
&&\nonumber\\
&=& -\tilde{\mathcal{K}}\tilde{\mathcal{T}}\tilde{h}_{ab}
\tilde{\nabla}^b \left[ \ln \left( 
\tilde{\mathcal{T}}\sqrt{-\tilde{g}_{00}} \right)\right] = 
0\,,\label{conformal_eckart_thermal_equilibrium} \end{eqnarray} where 
Eq.~(\ref{conformal_four_acceleration_assumption}) was used.
 
The Einstein frame fluid being at thermal equilibrium implies that 
$\tilde{\nabla}^b \left[ \ln \left( 
\tilde{\mathcal{T}}\sqrt{-\tilde{g}_{00}}\right)\right]$ is parallel to 
$\tilde{u}^b$ and orthogonal to $\tilde{h}_{ab}$, there exists an 
integration constant $\tilde{\mathcal{T}}_0$ such that \be
    \tilde{\mathcal{T}}\sqrt{-\tilde{g}_{00}} = \tilde{\mathcal{T}}_0\,, 
    \label{tolman_ehrenfest_einstein_frame}
\ee 
a near identical match  with the Tolman-Ehrenfest 
criterion~(\ref{tolman_ehrenfest_criterion_gr}) of the Jordan frame.  
Dividing term to term Eqs.~(\ref{tolman_ehrenfest_einstein_frame}) 
and~(\ref{tolman_ehrenfest_criterion_gr}) gives 
\be 
\tilde{\mathcal{T}}\Omega\sqrt{-g_{00}} = 
\bigg(\frac{\tilde{\mathcal{T}}_0}{ 
\mathcal{T}_0}\bigg)\,\mathcal{T}\sqrt{-g_{00}} 
\ee 
and, absorbing the 
ratio of the constants $ \mathcal{T}_0$ and $\tilde{\mathcal{T}}_0$ into $\Omega$ 
via a coordinate redefinition, this simplifies to  
\be \tilde{\mathcal{T}} = \frac { 
\mathcal{T} }{\Omega} \,. \label{temperature_conformal_relationship} 
\ee 
This relation has been demonstrated for more general conformally static 
spacetimes in the context of GR \cite{Faraoni:2023gqg}. The most physical 
example is the scaling of the cosmic microwave background with the scale 
factor of the Friedmann-Lema\^itre-Robertson-Walker (FLRW) universe in 
which it lives \cite{Faraoni:2023gqg}. Restrict, for simplicity, to a 
spatially flat FLRW 
universe with line element \begin{equation} ds^2=- d t^2 + a^2(t)\left( 
dx^2+ dy^2+ dz^2 \right) \end{equation} in comoving coordinates 
$\big(t,x,y,z\big)$. The use of the conformal time $\eta$, defined by $ dt 
= a \, d\eta$, makes the conformal flatness of this geometry explicit: 
\begin{equation} ds^2=a^2(\eta)\big(- d\eta^2 + dx^2+dy^2+dz^2\big)\,, 
\end{equation} with conformal factor $\Omega$ equal to the scale factor 
$a(\eta)$. Long after equipartition, the cosmic microwave background 
becomes a test fluid in local thermal equilibrium that evolves decoupled 
from other forms of matter. To maintain the Planck spectral energy density 
distribution \be
   u\left( \nu, {\cal T} \right) = \frac{8\pi h\nu^3}{c^3} \, 
\frac{1}{\mbox{e}^{\frac{h\nu}{k_B {\cal T} }} -1} \ee (where $h$ is the 
Planck constant, $k_B$ is the Boltzmann constant, $\nu $ is the photon 
frequency, and we restore the speed of light $c$), the temperature ${\cal 
T}$ of the  cosmic microwave background must scale as $ {\cal T} 
\propto 1/a $ 
because the proper wavelength $\lambda$ scales with the scale factor $a$ 
like all physical lengths, i.e., $h\nu = hc / \lambda \propto 1/a $.  The 
scaling $ {\cal T} \propto 1/a $ is nothing but the scaling $\tilde{ {\cal 
T}} = {\cal T}/ \Omega $ derived here, but in a very different context. 
Here we discuss thermal equilibrium (restricted to static spacetimes) in 
the context of scalar-tensor gravity, whereas Ref.~\cite{Faraoni:2023gqg} 
treats conformally static time-dependent spacetimes in GR.

 The transformation of the four-acceleration $\tilde{\dot{u}}^a $ 
under a conformal transformation $g_{ab} \to \tilde{g}_{ab} = \Omega^2 
g_{ab}$, 
\be 
\tilde{\dot{u}}^a = \frac{\dot{u}^a }{\Omega^2}  +\frac{ \nabla^a 
\Omega }{\Omega^3} + u^a \left( \frac{u^c \nabla_c \Omega }{\Omega} 
\right) \,, \label{accelerationtilde}
\ee
is derived in Appendix~\ref{Appendix:A}. 

Let us examine the tranformation properties of the stress-energy 
tensor of 
the dissipative test fluid. The variational definition of the 
Einstein frame stress-energy tensor 
 \begin{eqnarray} 
\tilde{T}_{ab} &=& \frac{-2}{ \sqrt{ 
-\tilde{g} } } \, \frac{ \delta \left( \sqrt{ -\tilde{g} } \, \tilde{ 
{\cal  L}}^\mathrm{(m)} \right) }{\delta \tilde{g}^{ab} }\,,
\end{eqnarray}
where  $ \tilde{ {\cal 
L}}^\mathrm{(m)}$ is the matter Lagrangian density  in the 
Einstein frame, implies that 
$ \tilde{ {\cal L}}^\mathrm{(m)}=\Omega^{-4} {\cal L}^\mathrm{(m)}$,  
and 
$ \sqrt{-\tilde{g}} \, \tilde{ {\cal L}}^\mathrm{(m)}= \sqrt{-g} \, {\cal 
L}^\mathrm{(m)}$, and (e.g., \cite{Faraoni:2004pi}) 
\be
    \tilde{T}^{ab} = \Omega^{-6}\,T^{ab}\,, \quad\quad \tilde{T}_{ab} = 
\Omega^{-2} \,T_{ab}\,,
    \label{conformal_transformation_stress_energy_tensor} 
\ee 

Using this fact and $\tilde{u}_a = 
\Omega u_a$, $ \tilde{h}_{ab}=\Omega^2 h_{ab}$, one writes
\begin{eqnarray}
\tilde{T}_{ab} &=& \frac{T_{ab}}{\Omega^2} = 
\frac{\rho u_a u_b}{\Omega^2} +
\frac{ P h_{ab}}{\Omega^2} +
\frac{ \pi_{ab}}{\Omega^2} +
q_a \, \frac{ u_b}{\Omega^2} +
q_b \, \frac{ u_a}{\Omega^2} \nonumber\\
&&\nonumber\\
&=& \frac{\rho}{\Omega^2} \, \frac{\tilde{u}_a}{\Omega} \, 
\frac{\tilde{u}_b}{\Omega} 
+ \frac{ P \tilde{h}_{ab}}{\Omega^4} 
+ \frac{ \pi_{ab}}{\Omega^2} +
\frac{ q_a}{\Omega^2}  \, \frac{ \tilde{u}_b}{\Omega} +
\frac{ q_b}{\Omega^2}  \, \frac{ \tilde{u}_a}{\Omega}
 \nonumber\\
&&\nonumber\\
&=& \tilde{\rho} \tilde{u}_a \tilde{u}_b + 
\tilde{P} \tilde{h}_{ab} +
 \tilde{\pi}_{ab} +
\tilde{q}_a \tilde{u}_b + \tilde{q}_b \tilde{u}_a \,,
\end{eqnarray}
where 
\begin{eqnarray}
\tilde{\rho} &=& \Omega^{-4} \rho \,,\\
&&\nonumber\\ 
\tilde{P} & = & \Omega^{-4} P \,,\\
&&\nonumber\\  
\tilde{\pi}_{ab} & = & \Omega^{-2} \pi_{ab} \,,\\
&&\nonumber\\  
\tilde{q}_a & = & \Omega^{-3} q_a \,.\label{macheczz}
\end{eqnarray}
The transformation properties of $\rho$ and $P$ for perfect and imperfect 
fluids are well known in the literature 
\cite{Faraoni:2004pi,Faraoni:2023gqg}. 
The scaling of the heat flux density $\tilde{q}_a=\Omega^{-3} q_a $  shows 
that thermal equilibrium in the Jordan frame (i.e., $q_a=0$) is 
equivalent to thermal equilibrium in the  
Einstein frame ($\tilde{q}_a=0$). All these rescalings are consistent with 
Dicke's \cite{Dicke:1961gz} dimensional arguments.

Let us show now that the Eckart constitutive relation in the Jordan frame 
follows from that in the Einstein frame.  Using the transformation 
properties $\tilde{ {\cal T}} = {\cal T}/\Omega$, $\tilde{q}_a = 
\Omega^{-3} q_a $ and Eq.~(\ref{accelerationtilde}), the Einstein frame 
relation
\be
\tilde{q}_a = -\tilde{K} \tilde{h}_{ab} \left( \tilde{\nabla}^b \tilde{ 
\cal{T} } + \tilde{ \cal{T} } \tilde{\dot{u} }^b \right)
\ee
yields
\begin{eqnarray}
\Omega^{-3} q_a &=&  -\tilde{K} \Omega^2 h_{ab} \left[ 
\Omega^{-2} g^{bc} \nabla_c \left( \frac{ \cal{T} }{\Omega} \right) 
\right.\nonumber\\
&&\nonumber\\
&\, & \left.  + 
\frac{  \cal{T} }{\Omega} \left(  \frac{ \dot{u}^b }{ \Omega^2} 
+\frac{\nabla^b \Omega}{\Omega^3} + 
\frac{u^c \nabla_c\Omega }{\Omega^3}   \right) \right] \nonumber\\
&&\nonumber\\
&=& -\tilde{K} \Omega^2 h_{ab} \left( \frac{\nabla^b {\cal T} }{\Omega^3} 
-\frac{ {\cal T} \nabla^b \Omega}{\Omega^4} + 
\frac{ {\cal T} \dot{u}^b }{\Omega^3} + 
\frac{ {\cal T} \nabla^b \Omega}{\Omega^4}  \right) \nonumber\\
&&\nonumber\\
&=& - \frac{ \tilde{K} \Omega^2 }{\Omega^3} \, h_{ab} \left( \nabla^b 
{\cal T} +{\cal T} \dot{u}^b \right) \,,
\end{eqnarray}
which reproduces the Eckart generalization of Fourier's law with 
$\tilde{K}=\Omega^{-2} K $.  By  repeating these steps in reverse, one 
obtains Eckart's law in the Einstein frame from that in the Jordan frame. 
Therefore, Eckart's law holds in the Einstein frame if and only if it 
holds in the Jordan frame.

\section{Stealth solutions} 
\label{sec:4} 

Let us consider now stealth solutions of the form $ g_{ab}=\eta_{ab}$, 
$\phi=\phi(t) $ (where $\eta_{ab}$ is the Minkowski metric). These 
solutions \cite{Ayon-Beato:2005yoq,Robinson:2006ib,Ayon-Beato:2004nzi,Demir:2006ed,Hassaine:2006gz,Takahashi:2020hso,Faraoni:2022jyd} are not 
possible in GR and are typical of scalar-tensor gravity. For these exotic 
stealth solutions, the effective stress-energy tensor $T_{ab}^{(\phi)}$ of 
the scalar field vanishes due to cancellations between its terms. 
Therefore, the Brans-Dicke-like scalar $\phi$ does not gravitate, but the 
effective gravitational coupling $G_\mathrm{eff}=1/\phi$ felt by test 
matter changes in time.

The derivation of the  criterion of thermal 
equilibrium in 
GR and in scalar-tensor gravity becomes invalid 
because, for stealth solutions, the Einstein frame metric 
$\tilde{g}_{ab}=\phi(t) \eta_{ab}$ is now time-dependent and the Buchdahl 
relation~(\ref{four_acceleration_general}) does not apply. However, the 
Einstein frame metric $\tilde{g}_{ab}$ is special since it is conformally 
flat, and the discussion of Ref.~\cite{Faraoni:2023gqg} applies to such 
situations. We do not repeat that discussion here, but report the salient 
logical steps. Although the Buchdahl relation is invalid, in conformally 
flat geometries\footnote{The discussion of Ref.~\cite{Faraoni:2023gqg} is 
more general since it applies to {\em conformally static} geometries.} 
$\tilde{g}_{ab}=\Omega^2(t) \eta_{ab} $, its projection onto the 3-space 
orthogonal to the fluid four-velocity $\tilde{u}^a$ still holds, that is, 
\be 
\tilde{h}_{ab} \tilde{a}^b = \tilde{h}_{ab} \tilde{\nabla}^b \ln 
\sqrt{-\tilde{g}_{00} } \,. 
\ee 
This fact is sufficient to conclude again that $\tilde{ {\cal T}}={\cal 
T}/\Omega$. Examples of this relation are the scaling of the temperature 
of the cosmic microwave background ${\cal T}\sim 1/a$ with the scale 
factor of the FLRW universe discussed in Sec.~\ref{sec:3}, and the scaling 
of the Hawking temperature of dynamical black holes embedded in a FLRW 
universe \cite{Faraoni:2023gqg}. Hence, for stealth solutions, the Jordan 
frame result is still valid and ${\cal T}$ does not depend on time.  
Observers comoving with the (Jordan frame) test fluid would find 
themselves in Minkowski space and would observe the fluid in thermal 
equilibrium with no heat flux and a temperature uniform and constant in 
time.  Stealth solutions of the field equations of scalar-tensor gravity 
are undoubtedly rather exotic and we will not discuss them further.

\section{Conclusions} 
\label{sec:5} 

We have extended the Tolman-Ehrenfest criterion for thermal equilibrium in 
GR to scalar-tensor gravity.   
The valid GR criterion~(\ref{tolman_ehrenfest_criterion_gr}) also 
holds  in 
Jordan frame and Einstein frame scalar-tensor gravity. One would expect 
the strength of the 
gravitational coupling $G_\mathrm{eff}=1/\phi$ to   
affect thermal 
equilibrium, but this is not the case. This fact is surprising 
because the essence of thermal equilibrium in GR (expressed by the 
Tolman criterion~(\ref{tolman_ehrenfest_criterion_gr})) is that heat is a 
form of mass-energy, therefore it sinks in a gravitational field. While, 
in GR, the gravitational coupling strength $G$ is a true constant, in 
scalar-tensor gravity it is not, thus one expects heat to  sink more where 
the 
gravitational coupling strength $G_\mathrm{eff}$ is higher and $\phi$ 
to   
appear in the condition for thermal equilibrium in the Jordan frame of 
scalar-tensor gravity. However, neither the field 
equations nor the conservation of the fluid's stress-energy tensor 
(which fails in the Einstein frame \cite{Dicke:1961gz}) are used in 
the derivation of the Tolman-Ehrenfest criterion.

 It is now 
tempting to consider extending the Tolman-Ehrenfest criterion to Horndeski 
and DHOST gravity,  but the transformation to a sort of analogous 
Einstein 
frame becomes much more complicated 
\cite{Deffayet:2009wt,Deffayet:2009mn,Deffayet:2011gz,Gleyzes:2014dya,Gleyzes:2014qga,Langlois:2015cwa,Langlois:2015skt,BenAchour:2016cay,Crisostomi:2016czh,BenAchour:2016fzp,Crisostomi:2017aim, Faraoni:2023hin,Langlois:2018dxi,Langlois:2017mdk,Creminelli:2018xsv,Langlois:2017dyl,Langlois:2018dxi,Kobayashi:2011nu}. 
It is a   
disformal transformation involving  first (and possibly higher order 
\cite{Takahashi:2021ttd,Takahashi:2023vva}) derivatives of the scalar 
field, which transforms the 
theory 
to a much more complicated and analogous Einstein frame where the role of 
scaling units (or its generalization) is completely unclear.  Since the 
field equations are not used, 
however, one 
expects the Tolman-Ehrenfest criterion to hold in any theory of gravity.    
The result presented here will be useful in 
studies of the cosmic microwave background and of Hawking radiation in 
scalar-tensor gravity, in the same way that the Tolman criterion is 
applied to these subjects in GR \cite{Faraoni:2023gqg,Miranda:2024xda}.

\bmhead{Acknowledgements}

V.~F. is grateful to Serena Giardino, Narayan Banerjee and Robert 
Vanderwee for a useful discussion.

\section*{Declarations}

\begin{itemize}
\item Funding: This work is supported, in part, by the 
Natural Sciences \& Engineering Research Council of Canada (grant 
2023-03234 to V.~F.) and by a Bishop's University Graduate Entrance 
Scholarship (N.~K.).

\item Conflict of interest: the authors declare no conflict of interest.

\item Ethics approval and consent to participate: not applicable.

\item Consent for publication: not applicable.

\item Data availability: due to the theoretical nature of this work, no 
data have been used.

\item Materials availability:  not applicable.

\item Code availability: not applicable.

\item Author contribution: both authors contributed equally to this work.

\end{itemize}

%\bigskip
%\bigskip\noindent
%Springer journals and proceedings: 
%\url{https://www.springer.com/gp/editorial-policies}
%
%\bigskip\noindent
%Nature Portfolio journals: 
%\url{https://www.nature.com/nature-research/editorial-policies}
%
%\bigskip\noindent
%\textit{Scientific Reports}: 
%\url{https://www.nature.com/srep/journal-policies/editorial-policies}
%
%\bigskip\noindent
%BMC journals: 
%\url{https://www.biomedcentral.com/getpublished/editorial-policies}
%\end{flushleft}

\begin{appendices}
\section{Transformation of the four-acceleration and Buchdahl 
relation in the  Einstein frame} 
\label{Appendix:A} 

Here we  derive the transformation of the four-acceleration  and we  
prove the Buchdahl relation directly in the Einstein frame using 
the transformation properties of kinematic and geometric quantities under 
conformal rescalings. In fact, the four-velocities normalizations $g^{ab} 
u_a u_b=-1$ and $\tilde{g}^{ab} \tilde{u}_a \tilde{u}_b=-1$ imply that \be
    \tilde{u}^a = \frac{ u^a}{\Omega} \,,\quad\quad \tilde{u}_a = \Omega 
\, u_a\,. \ee The four-acceleration $\tilde{a}^b$ in the tilded world is 
calculated in terms of the four-acceleration $a^b$, obtaining 
\begin{eqnarray}
\tilde{\dot{u}}^b & \equiv & \tilde{u}^c \tilde{\nabla}_c \tilde{u}^b = 
\frac{ u^c}{\Omega} \, \tilde{\nabla}_c \left( \frac{ u^b}{\Omega} \right) 
\nonumber\\
&&\nonumber\\
&=& \frac{u^c}{\Omega^2} \left( \partial_c u^b +\tilde{\Gamma}^b_{cd} u^d 
\right) -u^b \left( \frac{  u^c\nabla_c \Omega}{\Omega^3} \right) 
\nonumber\\
&&\nonumber\\
&=& \frac{u^c}{\Omega^2} \left\{ \partial_c u^b + \left[ \Gamma^b_{cd} + 
\frac{1}{\Omega} 
\left( {\delta^b}_c \nabla_d \Omega + {\delta^b}_d \nabla_c\Omega - g_{dc} 
\nabla^b \Omega \right) \right] u^d \right\}
-u^b \left( \frac{  u^c\nabla_c \Omega}{\Omega^3} \right) \nonumber\\
&&\nonumber\\
&=& \frac{u^c}{\Omega^2} \left[
\left( \partial_c u^b +  \Gamma^b_{cd}u^d \right) 
+ \frac{1}{\Omega}
 \left( {\delta^b}_c \nabla_d \Omega + u^b  \nabla_c\Omega -  
u_c  \nabla^b \Omega \right)  \right]
-u^b \left( \frac{ u^c\nabla_c \Omega}{\Omega^3} \right) \nonumber\\
&&\nonumber\\
&=& \frac{1}{\Omega^2} \left[ u^c \nabla_c u^b 
+ \frac{1}{\Omega}
 \left( u^b u^d  \nabla_d \Omega 
+ u^b u^c  \nabla_c\Omega +  \nabla^b \Omega \right)  
-u^b \left( \frac{  u^c\nabla_c \Omega}{\Omega} \right)  \right]  
\nonumber\\
&&\nonumber\\
&=& \frac{ \dot{u}^c}{\Omega^2}  
+ \frac{\nabla^b \Omega}{\Omega^3 } + u^b \left( \frac{u^c \nabla_c 
\Omega}{\Omega^3} \right) \,.
\end{eqnarray} 
As a check, in the Einstein frame test particles subject only to gravity 
(which follow geodesics of the Jordan frame, $\dot{u}^a=0$)  do not follow 
geodesics of the metric $\tilde{g}_{ab}$ but are subject to the 
acceleration  $ \frac{\nabla^b \Omega}{\Omega^3 } + u^b \left( \frac{u^c 
\nabla_c 
\Omega}{\Omega^3} \right)$. The second term is parallel to the 
four-velocity $u^b$ and can be removed by going to an affine 
parametrization, but there remains the residual four-acceleration
\be
 \frac{\nabla^b \Omega}{\Omega^3 }=  
 \frac{\nabla^b \phi}{2 \phi^2 }= 
 \frac{ \tilde{\nabla}^b \phi}{2 \phi }=
 \tilde{\nabla}^b \left( \ln \sqrt{\phi} \right) \,,
\ee
which reproduces Eq.~(\ref{modified-geodesic-eq}). It is important to note 
that the explicit coupling of the fluid to $\tilde{\phi}$ (or to $\phi$) 
is already taken into account in the fluid's four-acceleration and does 
not need to be inserted by hand in the Einstein frame Eckart relation
$ 
\tilde{q}_a = - \tilde{K}\tilde{h}_{ab} \left( \tilde{\nabla}^b 
\tilde{ {\cal T} } +\tilde{ {\cal T} } \tilde{\dot{u} }^b \right) $. 
In other words, one does not need to add another heat flux term depending  
on $ \tilde{\nabla}_a \phi$ in this relation. The physical interpretation 
of 
this fact is that, although the fluid is coupled to $\phi$ (in fact, 
because of this coupling), it is always 
in thermal equilibrium with it and there is no heat flow between this 
fluid 
and $\phi$.

Now $\tilde{a}^b $ can be rewritten by 
using $a^b$ given by Eq.~(\ref{four_acceleration_static}),  which 
yields\footnote{Because of 
the pressure gradient $\nabla_aP$, the four-acceleration of the fluid at 
rest is not zero unless it is a dust: otherwise, this fluid would be 
freely-falling and could not be static. A pressure gradient or other force 
is necessary to keep this fluid static in the given gravitational field, 
preventing it from flowing.} \begin{eqnarray}
    \tilde{a}^b &=& \dot{\tilde{u}}^b = \Omega^{-2}\bigg[ 
\nabla^b\,\Big(\ln \sqrt{-g_{00}}\Big) + 
\frac{\nabla^b\,\Omega}{\Omega}\bigg] \nonumber\\
&&\nonumber\\
    &=&\tilde{\nabla}^b\,\Big( \ln \sqrt{-g_{00}} + \ln\Omega\Big) =
\tilde{\nabla}^b \left( \ln \sqrt{-\tilde{g}_{00}} \right)\,,
    \label{conformal_four_acceleration_simple} 
\end{eqnarray} 
where we 
used the fact that \be \tilde{\nabla}^bf = 
\tilde{g}^{bc}\,\tilde{\nabla}_cf = \Omega^{-2}\,g^{bc}\,\nabla_cf = 
\Omega^{-2}\nabla^b f \ee for any scalar function $f$.



\end{appendices}


\bibliography{tolmanst-bibliography}


\end{document}